\documentclass[%
reprint,
 amsmath,amssymb,
 aps,nofootinbib,
]{revtex4-1}
\usepackage{tabularx}
\newcommand{\mf}{\mathcal{}}

\usepackage{xcolor}
\usepackage[
  margin=2.5cm,
  includefoot,
  footskip=20pt,
]{geometry}
\usepackage[ colorlinks = true,
             linkcolor = blue,
             urlcolor  = blue,
             citecolor = blue,
             anchorcolor = green,
]{hyperref}

\usepackage{graphicx}
\usepackage{dcolumn}
\usepackage{bm}
\usepackage{physics}
\begin{document}

\title{All-optical Quantum State Engineering for Rotation-symmetric Bosonic States}
\author{Rajveer Nehra\textsuperscript{1}}
\email{rnehra@caltech.edu}
\author{Miller Eaton\textsuperscript{2}}
\email{me3nq@virginia.edu}
\author{Olivier Pfister\textsuperscript{2}}
\email{opfister@virginia.edu}
\author{Alireza Marandi\textsuperscript{1}}
\email{marandi@caltech.edu}
\thanks{* and $\dagger$ contributed equally to this work.}
\address{$^1$ Department of Electrical Engineering, California Institute of Technology, Pasadena, CA 91125, USA}
 \address{$^2$ Department of Physics, University of Virginia, Charlottesville, VA 22904, USA}
\date{\today}
\begin{abstract}
Continuous-variable quantum information processing through quantum optics offers a promising platform for building the next generation of scalable fault-tolerant information processors. To achieve quantum computational advantages and fault tolerance, non-Gaussian resources are essential. In this work, we propose and analyze a method to generate a variety of non-Gaussian states using coherent photon subtraction from a two-mode squeezed state followed by photon-number-resolving measurements. The proposed method offers a promising way to generate rotation-symmetric states conventionally used for quantum error correction with binomial codes and truncated Schr\"{o}dinger cat codes. We consider the deleterious effects of experimental imperfections such as detection inefficiencies and losses in the state engineering protocol. Our method can be readily implemented with current quantum photonic technologies.

\end{abstract}
\maketitle
\section{Introduction}
Quantum information processing (QIP) opens a new paradigm for next generation information processors,  offering significant advantages over classical analogues for various fields including computation, communication, metrology, and sensing. In the last couple of decades, QIP has been widely explored mostly over discrete variables with qubits on many physical platforms such as superconducting circuits, photonics, trapped ions, quantum dots, nuclear spins, and neutral atoms~\cite{clarke2008superconducting,o2009photonic,cirac1995quantum,PhysRevLett.83.4204, saffman2010quantum,vandersypen2001experimental}. Another universal paradigm for QIP makes use of continuous variables (CV) such as the amplitude- and phase-quadratures of the quantized electromagnetic field for information encoding and processing ~\cite{Lloyd_1999}. The key interest in CVQIP comes primarily from its unprecedented potential to generate a massively entangled scalable quantum states known as cluster states at room temperature~\cite{Chen2014_1, Larsen2019,Asavanant2019}.\\
While the large cluster states have been deterministically produced, they do not offer any exponential speedups with Gaussian gates and Gaussian measurements alone~\cite{Walls1994}. Therefore it is of paramount importance to include some non-Gaussian elements to achieve quantum computational advantages~\cite{Bartlett2002, Nick_MBQC_CV}. Non-Gaussianity can be achieved through measurements such as photon-number-resolved detection (PNRD)~\cite{Lita2008,Nehra_optica:19}, non-Gaussian gates such as the cubic phase gate~\cite{Ghose_Cubic,Yanagimoto:2020}, or by the inclusion of non-Gaussian resources such as binomial states~\cite{Binomial_code_2016}, Schr\"{o}dinger cat states~\cite{Ourjoumtsev2006_1, Eaton_2019}, and Gottesman-Kitaev-Preskill (GKP) state~\cite{Gottesman2001}. In optical systems, a deterministic implementation of such non-Gaussian gates and non-Gaussian states is experimentally challenging due to weak optical nonlinearities. However, these non-Gaussian elements can be probabilistically realized using Gaussian resources (squeezed states and linear optics) with non-Gaussian measurements such as PNRD. Various schemes based on photon- addition, subtraction, parity projection, and catalysis for generating quantum states with non-Gaussian Wigner functions have been proposed and demonstrated~\cite{Dakna1998,Ourjoumtsev2006_1, Eaton_2019, Zavatta2004, thekkadath2020engineering, takase2021generation}. \\
Not only are non-Gaussian states required for exponential speedup, they are also crucial for quantum error correction to achieve fault-tolerance in quantum computation as Gaussian operations cannot protect against Gaussian errors~\cite{Niset2009}. A particular instance of useful non-Gaussianity is the GKP state, which encodes discrete quantum information in a CV Hilbert space, and thereby can be used for qubit based computation~\cite{Gottesman2001,tzitrin2020progress}. The ideal form of GKP states requires infinite energy, but universality can still be achieve with approximated GKP states and Gaussian operations~\cite{baragiola2019}. While quantum error correction (QEC) with GKP states has recently been experimentally implemented in superconducting and trapped ion systems~\cite{campagne2020quantum,fluhmann2019encoding}, an all-optical means to generate GKP states has remained a challenge despite the existence of several proposals~\cite{Goltsman2001,Vasconcelos2010,Weigand2018,etesse2014proposal,Eaton2019, shi2019fault, fabre2020generation, hastrup2021measurement}. 

Another promising avenue to exploit non-Gaussianity for CVQIP include quantum states that are rotationally symmetric in phase space in analog to translation phase space symmetry in GKP states. Error correcting codes designed with such states take advantage of the fact that states with $K$-fold rotational symmetry have decompositions over the Fock-basis with $K$-periodic spacing, making it easier to ascertain when photon loss and dephasing errors occur. Such codes include binomial codes and cat codes, and Pegg-Barnett codes~\cite{grimsmo2020,Binomial_code_2016, Bergmann2016}. While binomial and cat codes have been recently demonstrated in superconducting quantum circuits~\cite{reinhold2020error,hu2019quantum}, their implementations in optical domain have remained elusive. \\
In this work, we introduce an all-optical method to generate rotationally symmetric states with 2-fold and 4-fold-symmetry, in particular, binomial code states and truncated cat code states. Our method requires resources as minimal as a two-mode squeezed vacuum (TMSV) state, easily accessible linear optics, and PNRD at low photon numbers. Our proposal requires squeezing levels below what has already been demonstrated~\cite{Vahlbruch2016}, and low PNR detection which is now feasible due to advances in highly efficient transition-edge sensors~\cite{Lita2008} and number-resolving superconducting nanowire single-photon detectors detectors~\cite{cahall2017multi, nicolich2019, endo2021quantum}, and time-multiplexed detection schemes~\cite{Achilles_2003}.\\
We show that by tuning the initial available resource squeezing and post-selecting on desired the PNR outcomes, our method can generate a variety of states in addition to exact binomial states and approximated cat states with the generation rates in the KHz-MHz depending on the detection scheme using state-of-the-art PNR detectors with high count rates. Moreover, these rotationally symmetric states can be exactly produced as opposed to GKP states, which are only approximated due to available finite squeezing limit. \\
Our paper is structured as follows. In Section~\ref{sec:code}, we provide an overview of bosonic quantum error correction with binomial codes. Section~\ref{sec:I} discusses the general framework of our method to generate a various rotation-symmetric non-Gaussian states. In Section~\ref{sec:PNR_Engineering}, we focus on generating binomial and cat-like codes in the ideal case as well as in the presence of experimental imperfections. In Section~\ref{sec:imper_bin}, we derive the conditions for desired detection efficiency for faithful error correction. 
We present an architectural analysis in Section~\ref{sec:arch} and we conclude and offer an outlook in Section~\ref{sec:conclusions}. 
\section{Rotation-symmetric Bosonic Codes: Recap}\label{sec:code}
A detailed discussion on rotationally symmetric quantum error-correcting codes can be found in Refs.~\cite{chuang1997bosonic, Michael2016,li2017cat,grimsmo2020}; here we briefly summerize the main concepts. The rotation-symmetric bosonic codes include the bosonic encodings that remain invariant under discrete rotations in phase space in the same manner as GKP encodings are invariant under phase space translations. Some representative examples of rotation-symmetric bosonic codes are binomial codes and cat codes, which have been recently demonstrated in circuit quantum electrodynamics (cQED) architectures~\cite{michael2016new,albert2018performance,hu2019quantum}. 
The logical code words are stabilized by the photon-number super parity operator defined as
\begin{equation}
    \hat{\Pi}_{K} :=e^{\frac{i2\pi\hat{n}}{K}},
\end{equation}
where $\hat{n} = \hat{a}^\dagger \hat{a}$ is the photon-number operator. Therefore, in order to satisfy $\hat{\Pi}_{K}|\psi\rangle = |\psi\rangle$, the state $|\psi\rangle$ must have support on every K-th Fock basis.

For a given error set $\mathcal{E} = \{\hat{E}_1, \hat{E}_2,...,\hat{E}_l\}$, a necessary and sufficient criterion for a faithful error correction is known as the Knill-Laflamme condition, mathematically defined as:
\begin{equation}
    \hat{P}_\mathcal{C}\hat{E}^\dagger_l \hat{E}_m\hat{P}_\mathcal{C} = \alpha_{l,m}\hat{P}_\mathcal{C}; \hspace{ 5mm}\forall\hspace{2mm} \hat{E}_l, \hat{E}_m \in \mathcal{E},
    \label{eq:KL}
\end{equation}
where $\hat{P}_\mathcal{C}$ is the projector defined over the code space $\mathcal{C}$ and  $\alpha_{l,m}$ are entries of a Hermitian matrix~\cite{knill1997}. For example,  let's consider the (4-fold symmetric) binomial code words 
\begin{equation}
    {|0_L\rangle}=\frac12\ket{0}+\frac{\sqrt{3}}2\ket{4}, \hspace{4mm} {|1_L\rangle}=\frac{\sqrt{3}}2\ket{2}+\frac12\ket{6},
    \label{eq:binom_codestates}
\end{equation}
whose amplitudes are the square roots of binomial coefficients. One can easily see that these code words are able to detect and faithfully correct the errors given by error set $\mathcal{E} = \{\mathbb{I}, \hat{a}, \hat{a}^\dagger \hat{a}\}$ as per Knill-Laflamme condition. 

Physically, the Knill-Laflamme conditions 
ensure that the probability for a single-photon loss or a dephasing error to occur are the same for each code word, making it impossible for the environment to distinguish between the logical basis states $\ket{0_L}$ and $\ket{1_L}$. This preserves the encoded quantum information (the state's amplitudes in the logical basis), as it is not 
deformed as a result of the error and can thus be recovered by means of unitary operations.  To understand this further, let us consider a single-photon loss error for quantum state $\ket{\psi}$ encoded using the code words defined in \ref{eq:binom_codestates}. The logical state $|\psi\rangle = \alpha |0_L\rangle + \beta\ket{1_L}$ is an even photon-number parity state which transforms to odd parity state $\ket{\psi'}= \sqrt{3}\alpha\ket{3} + \sqrt{3/2}\beta(\ket{1}+\ket{5})$ under a single-photon loss error. Thus we get: 
\begin{equation}
   \hat{a}\ket{\psi}\rightarrow\ket{\psi'} =  \alpha\ket{\Tilde{0}_L} + \beta\ket{\Tilde{1}_L}, 
\end{equation}
where $\ket{\Tilde{0}_L} = \ket{3}$ and  $\ket{\Tilde{1}_L} = 1/\sqrt{2}(\ket{1}+\ket{5})$ are the error words, which are orthogonal to the original code words. We note that the quantum information encoded in the complex amplitudes $\alpha$ and $\beta$ is preserved and can be faithfully recovered by mapping the error words $\{\ket{\Tilde{0}_L}, \ket{\Tilde{1}_L}\}$ to code words $\{\ket{0_L}, \ket{1_L}\}$ by means of unitary operations~\cite{Michael2016}. Since the single-photon loss changes the photon-number parity of the original state from even to odd, the parity measurements can be used as an error syndrome measurement. In optics, error syndrome measurements can be performed using highly efficient photon-number-resolving transition-edge sensors~\cite{Lita2008,Nehra_optica:19, morais2020precisely}.

In general, the photon loss process can be viewed more precisely by subjecting a quantum state to a completely-positive and trace-preserving (CPTP) bosonic channel with non-unity transmission. In this case, the effect of the channel is given by the Kraus operator-sum representation formulated as
\begin{equation}
    \rho'=\mathcal{L}(\rho_0)=\sum_{k=0}^\infty \hat{E_k} \rho_0 \hat{E_k}^\dag, 
    \label{eq:Kraus}
\end{equation}
where $\rho'$ and $\rho_0$ are the output and input states, respectively, and to ensure the CPTP channel, one needs ${\sum_{k=0}^{\infty}E_k^\dagger E_k = \mathbb{I}}$. 
In this work, we treat loss as an amplitude-damping Gaussian channel, which can be modeled as an optical mode traveling a distance $L$ through a medium with loss coefficient $\alpha$ dB/cm. The Kraus operator associated with losing $k$ photons is
\begin{equation}
    \hat{E_k}=\sqrt{\frac{(1-e^{-{\gamma}})^k}{k!}}e^{-\tfrac{1}{2}\gamma \hat{a}^\dag \hat{a}}\hat{a}^k,
    \label{eq:kth_kraus}
\end{equation}
where we define $\gamma = \alpha L$. A $k$-photon loss to $\rho$ occurs with the probability of
\begin{equation}
    P_k=\langle \hat{E_k}^\dag \hat{E_k}\rangle=\frac{\gamma^k}{k!}\text{Tr}[\hat{a}^k\rho\, \hat{a}^\dag{}^k]+\mf{O[\gamma^{k+1}]}.
  \end{equation}
 With this in mind, QEC codes with the ability to correct up to $k$-photon losses can be reframed as codes that correct operators $E_k$ up to $k$-th order in $\gamma$. As a result, the code words we propose to create in Eq.~\ref{eq:binom_codestates} have 4-fold-symmetry in phase space with the ability to correct single photon losses, making it effective up to first order in $\gamma$ and corrects for $E_0$ and $E_1$ corresponding to `no-jump' and `single-jump' errors, respectively. 
\section{Proposed Method: Analytical Model}\label{sec:I}
\begin{figure*}[!htb]
    \centering
 \includegraphics[width = 1\textwidth]{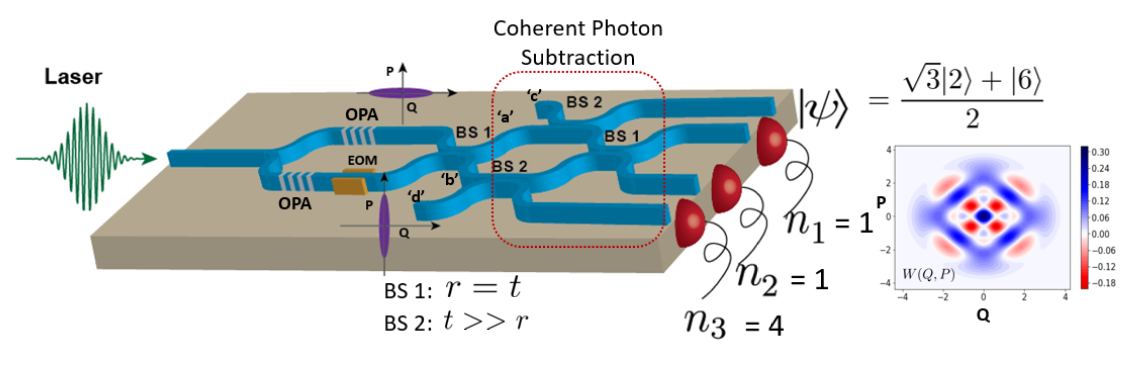}
  \caption{The proposed scheme for generating rotation-symmetric error-correcting codes. A two-mode squeezed vacuum (TMSV)  state is generated by combining two orthogonal single-mode squeezed vacuum (SMSV) states at the first balanced beamsplitter (BS 1). Photons are then subtracted from each modes of the TMSV state using highly unbalanced beamsplitters (BS 2). The subtracted photons are interfered at the second balanced beamsplitter followed by PNR measurements on three output modes, which prepares the fourth mode in the desired code for a certain measurement combination of $n_1, n_2$, and $n_3$. The dotted red box represents the coherent photon subtraction process. A particular case shown here considers $[n_1, n_2, n_3]$ = $[1,1,4]$, resulting in $\ket{\psi}\propto \ket{2}+\ket{6}$ with 4-fold rotational symmetry  as evident in the displayed Wigner function. OPA: Optical parametric amplifier. BS 1: Balanced beamsplitter. BS 2: Unbalanced beamsplitter for photon subtraction. EOM: Electro-optic modulator. }
    \label{fig:schematic}
    \end{figure*}

In this section, we detail the proposed method displayed in Fig.~\ref{fig:schematic}. Our method starts by preparing by a two-mode squeezed vacuum (TMSV) state by interfering two orthogonal single-mode squeezed vacuum (SMSV) states produced by  optical parametric amplifiers (OPAs) at the first balanced beamsplitter labeled as BS 1 in Fig.~\ref{fig:schematic}.
This is followed by two highly unbalanced beamsplitters (BS 2) used for photon subtractions from each mode of the TMSV state. Next, a balanced beamsplitter (BS 1) is placed to interfere the subtracted photons in order to erase the information about from which mode the subtracted photons came from. As a result, this combination of photon subtractions and the balanced interference allows one to coherently subtract photons from the TMSV state. Finally, photon-number-resolving (PNR) measurements are performed on three output modes prepares the desired state  $|\psi\rangle$ in the fourth mode for a certain combination of PNR measurement outcomes. \\
Interfering two SMSV states at a balanced beamsplitter is equivalent to preparing a TMSV state by action of the two-mode squeezed operator $S(z)_{ab}=e^{[z\hat{a}^\dag \hat{b}^\dag - z^*\hat{a} \hat{b}]}$ on two-mode vacuum, where $z=\mf{R}e^{i\phi}$ with real squeezing parameter $\mf{R}$ and phase $\phi$. In the photon-number basis, the TMSV state prepared in modes `a' and `b' after the first BS 1 is given as~\cite{Walls1994}
\begin{equation}
    |0,\mf{R}\rangle_{ab}=\frac{1}{\cosh{{R}}}\sum_{n=0}^\infty e^{in\phi}\tanh^nR|nn\rangle_{ab}, 
\end{equation}
After encountering the highly unbalanced beamsplitters (BS 2), the state interferes with vacuum modes `c' and `d', which leads to the four-mode state
\begin{equation}
     |\psi\rangle_{\text{abcd}} = \hat{U}_{ac}(\theta)\hat{U}_{bd}(\theta)|0,\mf{R}\rangle_{ab}|00\rangle_{cd}, 
     \label{eq:four_mode_state}
\end{equation}
where $\hat{U}_{ac}(\theta),\hat{U}_{bd}(\theta)$ are beamsplitter unitary operators where $(t, r) =  (\text{cos}\theta,\text{sin}\theta)$ are the transmission and reflections coefficients, respectively. We first consider the simplest case where the probability of subtracting more than one photon is negligible, i.e., at most one photon is subtracted from each mode. For single photon subtraction, we set both beamsplitters such that $t\rightarrow 1$ ($\theta<<1$) which allows one to approximate the unitary operators as 
\begin{align}
    \hat{U}_{ac}(\theta) &= \text{exp}[\theta(\hat{a}^\dagger \hat{c}- \hat{a}\hat{c}^\dagger)] \approx \mathbb{I} + \theta[\hat{a}^\dagger \hat{c}- \hat{a}\hat{c}^\dagger] + \mf{O}(\theta^2), \nonumber\\
       \hat{U}_{bd}(\theta) &= \text{exp}[\theta(\hat{b}^\dagger \hat{d}- \hat{b}\hat{d}^\dagger)] \approx \mathbb{I} + \theta[\hat{b}^\dagger \hat{d}- \hat{b}\hat{d}^\dagger] + \mf{O}(\theta^2),
       \label{eq:approx_BS_1}
\end{align}
where we consider $\mathcal{O}(\theta^2)$ terms negligible to ensure that mostly single photons are subtracted from each mode. 
From Eq.~\ref{eq:four_mode_state} and Eq.~\ref{eq:approx_BS_1}, we get the unnormalized four-mode state 
\begin{widetext}
\begin{equation}
    \hat{U}_{ac}(\theta)\hat{U}_{bd}(\theta)|0,\mf{R}\rangle_{ab}|00\rangle_{cd}\propto \sum_n^\infty e^{in\phi}\text{tanh}^nR[\mathbb{I}+\theta(\hat{a}^\dag \hat{c}-\hat{a}\hat{c}^\dag +\hat{b}^\dag \hat{d} -\hat{b}\hat{d}^\dag)]|nn\rangle_{ab}|00\rangle_{cd},\\
    \label{eq:combined_unitaries}
\end{equation}
Finally, the action of last balanced beamsplitter (BS 2) leads to the resulting state 
\begin{equation}
    |\psi\rangle_{\text{abcd}}\propto \sum_n^\infty e^{i n\phi}\text{tanh}^nR\bigg[|00\rangle_{cd} - \frac{\theta}{\sqrt{2}}\big[(\hat{a}-\hat{b})|10\rangle_{cd}+(\hat{a}+\hat{b})|01\rangle_{cd}\big]\bigg]|nn\rangle_{ab}.
    \label{eq:four_mode_output}
\end{equation}
\end{widetext}
We now perform the PNR measurements on three output modes as depicted in Fig.~\ref{fig:schematic}. Looking at the state in Eq.~\ref{eq:four_mode_output}, it can be immediately seen that for  $n_1 = n_2 = 0$, the initial TMSV state is unchanged as one expects. We now consider two cases of $n_1 = 1, n_2 = 0$ and $n_1 = 0, n_2 = 1$, i.e., only one of the two detectors detects a single photon. The two orthogonal output states in these two cases are
\begin{equation}
    |\psi\rangle^\pm_{ab} \propto \sum_{n}^\infty \sqrt{n}e^{in\phi}\text{tanh}^nR[\ket{n-1,n} \pm \ket{n,n-1}]\label{eq:simple_case},
\end{equation}
where the positive (negative) sign corresponds to [$n_1, n_2] = [1, 0]$ ([$n_1, n_2] = [0, 1]$).

It is worth mentioning that states like $\ket{\psi}^\pm$ have been shown to achieve Heisenberg-limit of phase measurements in quantum interferometry~\cite{Hofmann2006, Miller2021}. 
We now consider a third PNR detector placed in the path of mode `b' to detect $n_3$ photons. The third PNR measurement projects the $\rho^\pm$ to the pure state given by

\begin{align}
    |\psi\rangle^{\pm}_a =& \mathcal{N}[\sqrt{n_3c_{n_3,n_3}} |n_3 - 1\rangle \nonumber\\  &\pm \sqrt{(n_3+1)c_{n_3+1,n_3+1}} |n_3 + 1\rangle], \label{eq:final_state_1}
\end{align}
where 
\begin{equation}
   \mathcal{N} = \sqrt{n_3c_{n_3,n_3}+(n_3+1)c_{n_3+1, n_3+1}}
\end{equation} 

\begin{equation}
  c_{n,n'}= {e^{i\phi(n+n')}\tanh^{n+n'}{\mf{R}}}/{\cosh^2{\mf{R}}}
\end{equation} 

From Eq.~\ref{eq:final_state_1}, one can see that for $n_3$ = odd (even), the final state has two consecutive even (odd) Fock components corresponding to $n_3-1$ and $n_3+1$.
We note that this particular task of coherent single-photon subtraction with highly unbalanced beamsplitters can be performed with click detectors by ensuring that the probability of reflecting more than one photon is vanishing. However, the third detector needs to be PNR detector. In Section~\ref{sec:PNR_Engineering}, we show that how one can generate rotation-symmetric states with 2-fold and 4-fold phase space symmetry in the Wigner function.\\
We now generalize the proposed method for arbitrary detection of $n_1$ and $n_2$ photons. In this case, to the leading order approximation, only the terms involving up to $\theta^{n_1+n_2}$ will contribute in the expansion of unitary operators of the photon subtracting beamsplitters (BS 2 in Fig.~\ref{fig:schematic}) in Eq.~\ref{eq:combined_unitaries}. As a result, the Taylor expansion can be approximated to
\begin{equation}
    \hat{U}_{ac}(\theta)\hat{U}_{bd}(\theta)\approx\sum_{k=0}^{N}\frac{\theta^{N}}{k!(N-k)!}(\hat{a}^\dag \hat{c} - \hat{a}\hat{c}^\dag)^{N-k}(\hat{b}^\dag \hat{d} - \hat{b}\hat{d}^\dag)^k,
    \label{eq:bs_approx}
\end{equation}
where $N=n_1+n_2$ is the total number of photons detected after the photon subtraction step. Since we have vacuum inputs to modes $c$ and $d$, and as a consequence of detecting $N$ photons in total, Eq.~\ref{eq:bs_approx} reduces to
\begin{equation}
    \hat{U}_{ac}(\theta)\hat{U}_{bd}(\theta)\approx\sum_{k=0}^{N}\frac{(-\theta)^{N}}{k!(N-k)!}(\hat{a}\hat{c}^\dag)^{N-k}(\hat{b}\hat{d}^\dag)^k.
    \label{eq:subtoperator_apprx}
\end{equation}
 This approximation can now be transformed by the second balanced beamsplitter, $\hat{U}_{cd}(\theta)$, into
\begin{widetext}
\begin{equation}
    \hat{U}_{cd}\hat{U}_{ac}(\theta)\hat{U}_{bd}(\theta)\hat{U}_{cd}^\dag\approx\bigg(\frac{-\theta}{\sqrt{2}}\bigg)^{N}\sum_{k=0}^{N}\frac{\hat{{a}}^{N-k}\hat{b}^k}{k!(N-k)!}(\hat{c}^\dag+\hat{d}^\dag)^{N-k}(-\hat{c}^\dag+\hat{d}^\dag)^k.
\end{equation}
The action of these beamsplitters to the TMS resource state and vacuum modes $c$ and $d$ followed by detecting $n_2$ and $n_1$ photons in the transformed modes $c$ and $d$, respectively, leads to the final state
\begin{equation}
    |\psi\rangle_{ab} = \hat{U}_{cd}\hat{U}_{ac}(\theta)\hat{U}_{bd}(\theta)\hat{U}_{cd}^\dag| 0,R\rangle_{a,b}\otimes |0,0\rangle_{c,d} \propto\sum_{k=0}^{N}A_{n_1,n_2,k}\hat{a}^{N-k}\hat{b}^k|0,\mf{R}\rangle_{ab}, 
    \label{eq:tms_projected}
\end{equation}
where the coefficient $A_{n_1,n_2,k}$ is 

 \begin{equation}
     \label{eq:coefA}
   A_{n_1,n_2,k}=\sum_{i_{min}}^{i_{max}} \frac{(-1)^{i+k-n_>}\sqrt{n_1!n_2!}}{i!(i+k-n_>)!(n_>-i)!(N-i-k)!},
 \end{equation}
and the summation goes from $i_{min}=\text{Max}(0,n_>-k)$ to $i_{max}=\text{Min}(n_>,N-k)$, where $n_>=\text{Max}(n_1,n_2)$. The overall success probability of this combination of PNR measurements is
\begin{equation}
    P([n_1,n_2])=\Big(\frac{\theta}{\sqrt{2}}\Big)^{2N}\text{Tr}\Big[\sum_{k=0}^N\sum_{j=0}^NA_{n_1,n_2,k}A_{n_1,n_2,j}\hat{a}^{N-k}\hat{b}^k|0,\mf{R}\rangle_{ab}\langle0,\mf{R}|\hat{a}^{\dag^{N-j}}\hat{b}^{\dag^j}\Big].
\end{equation}
\end{widetext}
One salient point to note about the coherent photon subtraction is that the approximation of weakly reflective subtraction beamsplitters used in Eq.~\ref{eq:subtoperator_apprx} is not strictly necessary. In fact, as we show in appendix~\ref{appx:perfect_sub}, the full treatment of perfect coherent photon subtraction from a TMSV state with arbitrary beamsplitter parameter $\theta$ reduces to the same two-mode state in Eq.~\ref{eq:tms_projected}, but with a different effective squeezing parameter. If the initial experimental squeezing parameter is $\mf{R}$, then the nonzero beamsplitter reflectivity acts to replace $\mf{R}$ in Eq.~\ref{eq:tms_projected} with an effective squeezing parameter of $\mf{R}'=\text{tanh}^{-1}{(t^2\tanh{\mf{R}})}$, where $t^2=\cos^2{\theta}$ is the transmission of each subtraction beamsplitter. Later we show how one can utilize this feature of coherent photon subtraction to increase the success rates of preparing the desired states.
We now investigate the third PNR measurement on the coherently photon-subtracted state represented by Eq.~\ref{eq:tms_projected}. A measurement outcome of $n_3$ photons projects the two-mode state $|\psi\rangle_{a,b}$ to the following parity state
\begin{align}
     |\psi\rangle_{a}&=\mathcal{N}\sum_{k=k_{min}}^{N+n_3}A'_{n_1,n_2,k}|2k-N-n_3\rangle_{a},
    \label{eq:tms_projected_final}
\end{align}
where we have
\begin{equation}
   A'_{n_1,n_2,k}=A_{n_1,n_2,(k-n_3)}\frac{(e^{i\phi}\tanh{\mf{R}})^{k}k!}{\sqrt{(2k-N-n_3)!}},
   \label{eq:coefAprime}
\end{equation}
and $k_{min}=\text{Max}(n_3,\lceil\tfrac{1}{2}N+\tfrac12n_3\rceil)$ with $\lceil.\rceil$ being the ceiling function. The normalization coefficient is given by
\begin{equation}
    \mathcal{N}=\frac{\left(\tfrac{1}{2}\theta^2\right)^{N}}{n_3!(\cosh{\mf{R}})^2\sqrt{P_{\text{succ}}}}, 
\end{equation}
where $P_{\text{succ}}$ is the overall success probability of the measurement outcome configuration $[n_1, n_2, n_3]$ given as
\begin{equation}
    P_{\text{succ}}=\frac{\left(\tfrac{1}{2}\theta^2\right)^{N}}{n_3!(\cosh{\mf{R}})^2}\sum_{k = k_{min}}^{N+n_3}(A'_{n_1,n_2,k})^2.
    \label{eq:psucc_tot}
\end{equation}
This completes the generalized derivation for coherent photon subtraction scheme for generating rotation-symmetric error correcting codes. Examining Eq.~\ref{eq:tms_projected_final}, we see that in addition to the state having definite photon-number parity, the number of Fock components in the superposition is determined by the number of subtracted photons to reach a maximum of $N+1$ when $n_1\neq n_2$ and $n_3>N$. When $n_3<N$, the maximum number of components drops to $\lfloor \tfrac12 N+\tfrac12 n_3\rfloor+1$.
A particularly interesting case occurs when $n_1=n_2$. In this case the coefficient $A_{n_1,n_2,k}$ vanishes for $k = \text{odd}$ as seen below: 
\begin{widetext}
\begin{align}
    A_{n_1,n_2,k_{odd}}=\sum_{i=0}^{n_2-\frac{k-1}{2}} &(-1)^{n_2+k-i}\big[(2n_2-k-i)!(k+i-n_2)!i!(n_2-i)!)\big]^{-1} \nonumber\\
    &+(-1)^{-n_2-i}\big[i!(n_2-i)!(2n_2-k-i)!(k+i-n_2)!)\big]^{-1}=0.
\end{align}
\end{widetext}
Therefore, only the terms where $k = \text{even}$ contribute to the state in Eq.~\ref{eq:tms_projected_final} meaning that the final state has half as many Fock-basis components as the $n_1\neq n_2$ case, and each of these components is separated by integer multiples of 4. Indeed, this is a consequence of extended Hong-Ou-Mandel (HOM) type interference between the subtracted photons~\cite{HOM:1987}. Since we are considering only the case where $n_1 = n_2$, the subtracted photons from each mode of the TMSV can only occur in even numbers. The simplest case of such scenario is $n_1 = n_2 = 1$, which means both the subtracted photons came from either mode `a' or `b', as one expects in a typical HOM experiment. As we show later, this extended HOM interference property can be used to prepare rotationally-symmetric states which can be used in bosonic quantum error correction schemes~\cite{Binomial_code_2016,grimsmo2020}. \\ 
\section{Numerical Experiments}
\begin{figure*}
    \centering
    \includegraphics[width = 1\textwidth]{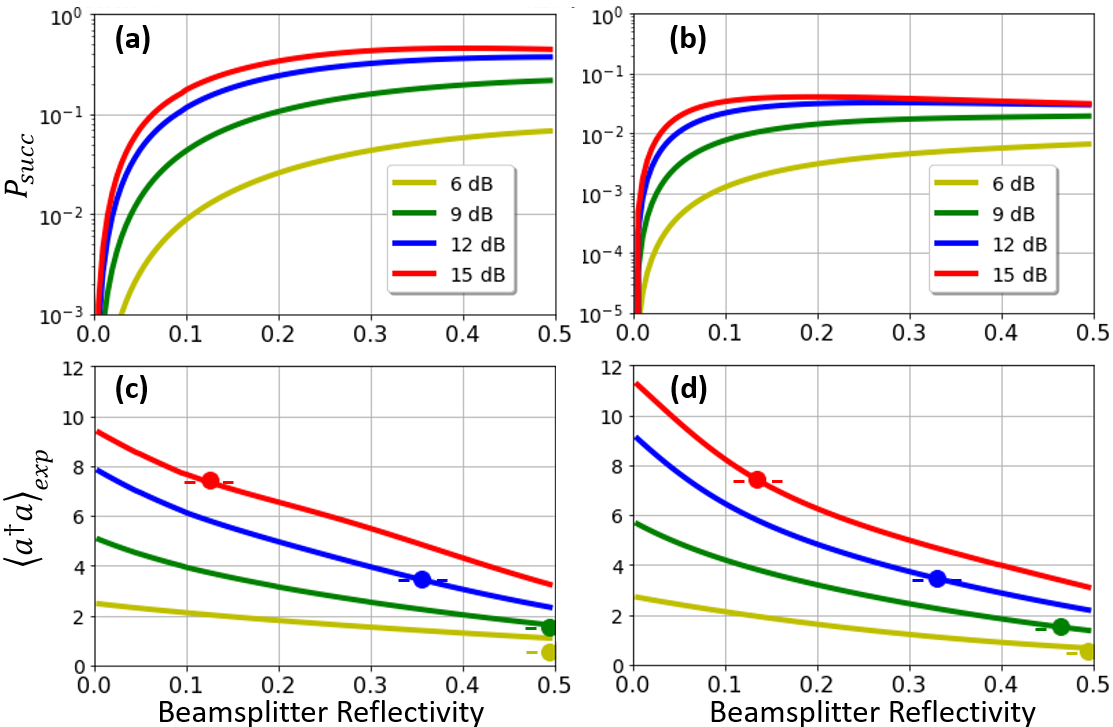}
\caption{(a) The probability to successfully generate any non-Fock parity state (2-fold or 4-fold-symmetry) as the subtraction beamsplitter reflectivity is changed for many values of 
initial two-mode squeezing. (b) Probability to successfully generate only 4-fold-symmetry states. (c),(d) The average of the mean-photon number of the successfully generated states. The large dots indicate the mean photon number of a single mode of input state at each respective squeezing. Regions on the curves to the left of the dots show it is likely to increase the average energy of the mode when the state is successfully generated.}
    \label{fig:all_statetypes}
\end{figure*}

In general for arbitrary $n_1, n_2,$ and $n_3$, we show in Fig.~\ref{fig:all_statetypes} how the probability to successfully generate any non-Fock state with either 2-or 4-fold phase space symmetry can be optimized by varying the subtraction beamsplitter (BS 2 in Fig.~\ref{fig:schematic}) reflectivies based on the initial resource squeezing level. Since changing the subtraction beamsplitter reflectivity has no effect on the form of the output state other than to reduce the original TMS state squeezing parameter to a smaller effective squeezing, this feature can be used as an external experimental knob to tune the mean output state energy and boost the generation rates. The `non-Fock' qualifier means we exclude generated states that are composed of only a single Fock-basis component. Thus, the probabilities plotted simply sum the probabilites to obtain any interesting superposition state of the desired symmetry for a given subtraction beamsplitter reflectivity at each initial squeezing.  Additionally, 4-fold rotation symmetric are also posses 2-fold symmetry, Fig.~\ref{fig:all_statetypes}a shows the probability to generate either type of states. In Fig.~\ref{fig:all_statetypes}b, we exclusively show the case of $n_1=n_2$ where 4-fold symmetric states are generated.

The plots demonstrate that there is a threshold value for the beamsplitter reflectivity beneath which the generation probability drops precipitously, but increasing the reflectivity beyond about $10\%$ ($r^2 = 0.1$) gives limited returns. Furthermore, in Fig.~\ref{fig:all_statetypes}c and \ref{fig:all_statetypes}d we show the expected mean photon number, $\langle a^\dag a \rangle_{\text{exp}}$ of the output state given the specified initial conditions. These values are obtained by averaging over the mean photon number of each output state weighted by the probability of obtaining that state while excluding the non-superposition states. As one can see, the value of $\langle a^\dag a \rangle_{exp}$ for the generated states only decreases as reflectivity increases, which intuitively follows considering more of the original TMS state energy is diverted toward the subtraction detectors. In all cases however, it is interesting to see that for all values of squeezing considered, the energy of the generated state can be increased beyond what was present in a single mode of the initial TMS state. The large dot corresponding to each squeezing curve in the plots shows the mean photon-number of each single mode of the initial resource state and thus the regions to the left of this point indicate that on average, successfully generating the desired symmetry state with the indicated squeezing and beamsplitter reflectivities will increase the mean photon-number of the output state. This effect is attributable to subtracting photons from two-mode squeezed vacuum, and has been noted previously in the context of quantum state interferometry~\cite{Carranza2012}. 
\section{State Engineering for Bosonic Error Correcting Codes}\label{sec:PNR_Engineering}
In this section, we show how the proposed scheme can be tuned to generate exact binomial- and truncated cat-like codes by choosing certain combinations of PNR measurements. We provide a detailed analysis on required resource squeezing, success probabilities, and how lossy PNR detection affects the quality of the generated bosonic codes.
\subsection{General two-component states}
As mentioned earlier in Sec.~\ref{sec:I}, simply by tuning the initial squeezing parameter and post-selecting for either  $n_1=1, n_2=0$ or $n_1=0, n_2=1$ produces any 2-fold symmetric two-component state of the form 
\begin{equation}
    \label{eq:pi_symm}
    \ket{\psi_\pi}\propto \ket{m-1}+\alpha\ket{m+1},
\end{equation}
 where the only requirement is that $|\alpha|<\sqrt{\tfrac{m+1}{m}}$. Similarly, detecting one photon in each of modes $c$ and $d$ ($n_1=n_2=1$) allows the production of arbitrary 4-fold symmetric two-component states of the form
\begin{equation}
    \label{eq:halfpi_symm}
    \ket{\psi_{\pi/2}}\propto \ket{m-2}+\beta\ket{m+2},
\end{equation}
where $|\beta|<\sqrt{\tfrac{(m+1)(m+2)}{m(m-1)}}$. In the above formulas, $\alpha$ and $\beta$ are complex numbers determined by the amplitude and phase of the squeezing parameter for the initial resource TMSV state. If the initial squeezing parameter is written in complex form as $z=\mf{R}e^{i\phi}$, then beginning with a state that satisfies
\begin{equation}
    e^{i\phi}\tanh{\mf{R}}=-\alpha\bigg(\frac{{m}}{{m+1}}\bigg)^{\frac{1}{2}}
    \end{equation}
and performing a PNR detection of $n_3=m$ after the coherent subtraction leads to the creation of the 2-fold-symmetry state with parameter $\alpha$ in Eq.\ref{eq:pi_symm}. In order to generate the 4-fold symmetric state in Eq.~\ref{eq:halfpi_symm}, the condition on the initial squeezing is
\begin{equation}
    e^{i\phi}\tanh{\mf{R}}=\sqrt{-\beta}\left(\frac{m(m-1)}{(m+1)(m+2)}\right)^{\frac{1}{4}},
    \label{eq:halfpi_cond}
\end{equation}
where again a PNR detection of $n_3=m$ must follow the coherent photon subtraction of the $n_1=n_2=1$ case. 
\subsection{Binomial Code State Generation}
\begin{figure*}[!tbh]
    \centering
    \includegraphics[width = 1\textwidth]{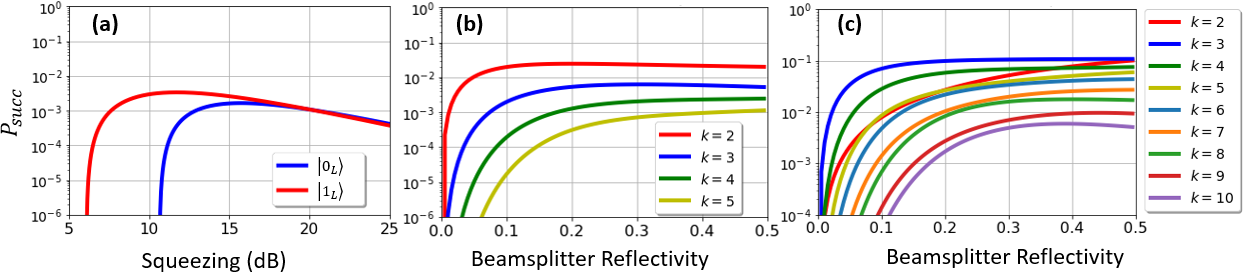}
\caption{(a) Probability to successfully generate the code states $\ket{0_L}$ and $\ket{1_L}$ as initial squeezing is increased. The subtraction beamsplitters are tuned to satisfy Eq.~\ref{eq:halfpi_cond}. (b) Probability to successfully generate any 4-fold symmetric state and (c) and 2-foldsymmetric state with $k$ Fock-basis components in the each superposition as subtraction beamsplitter reflectivity is increased given an initial squeezing of 12 dB.}
    \label{fig:fig_probsucc}
\end{figure*}

\begin{figure*}
    \centering
    \includegraphics[width = \textwidth]{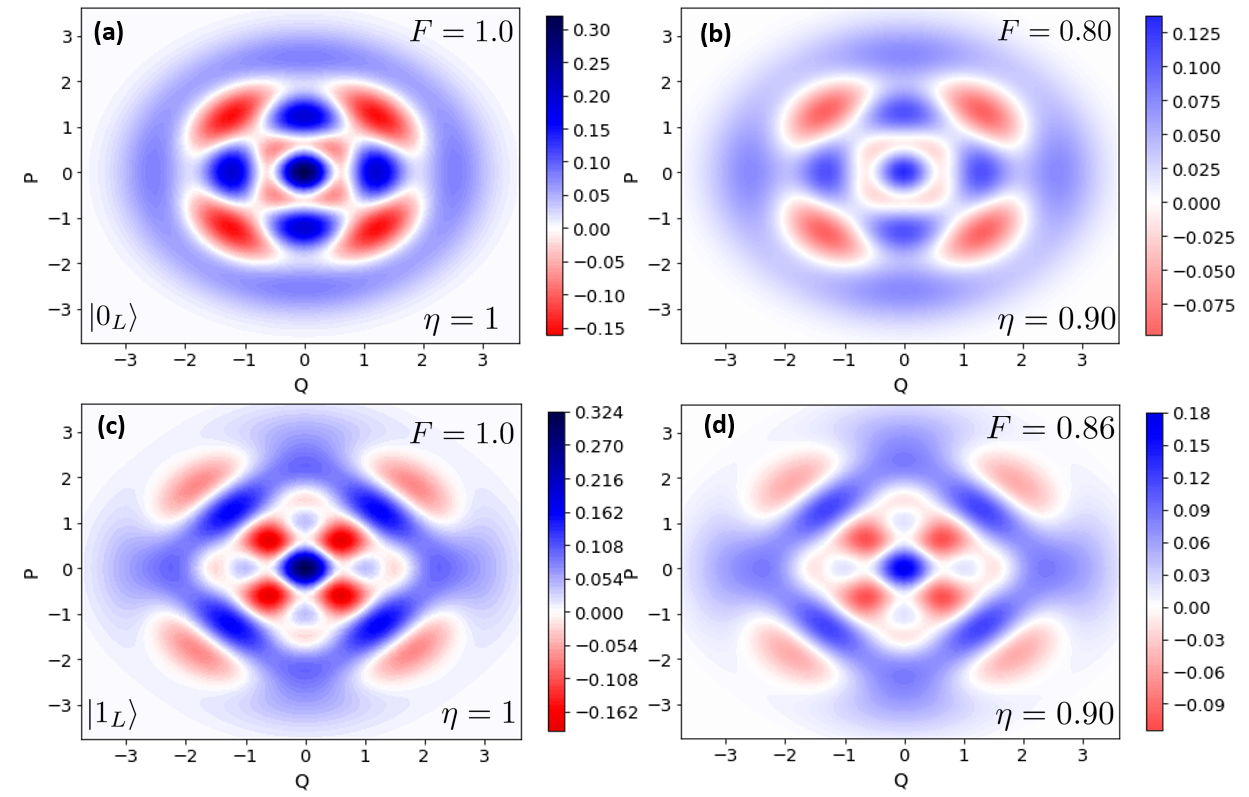}
    \caption{Wigner function visualization of binomial codewords. (a) and (c) are the ideal codewords for $\ket{0_L}$ and $\ket{1_L}$ when no losses are considered in our protocol. (b) and (d) are in the presence of 10\% losses in the PNR measurement of $n_3$ photons. We see that the features such as phase space interference and the negativity of the Wigner functions are preserved with 10\% losses. }
    \label{fig:binom_wig}
\end{figure*}

\begin{figure*}[!htb]
    \centering
    \includegraphics[width = \textwidth]{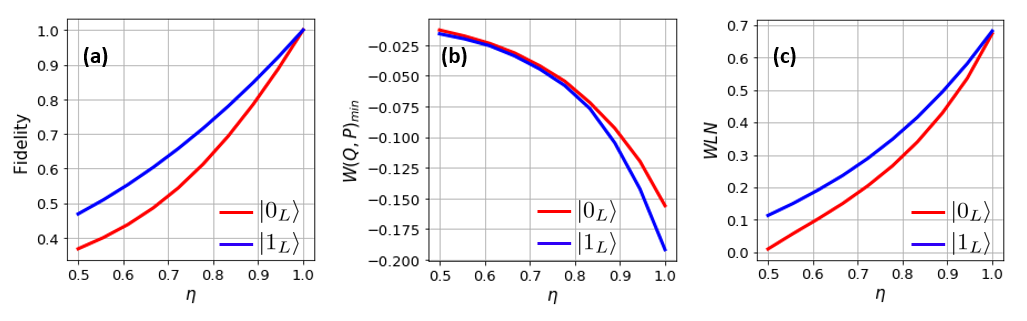}
    \caption{Effects of lossy PNR detection for $n_3$ photons. (a) State Fidelity with the ideal codewords. The minimum Wigner function amplitude and Wigner Log Negativity (WLN), (b) and (c), respectively. }
    \label{fig:wig_negativity}
\end{figure*}

We now consider two particular cases of 4-fold symmetric two-component states, in particular the binomial code words of Eq.~\ref{eq:binom_codestates}. These binomial code words ${|0_L\rangle}$ and ${|1_L\rangle}$ can be exactly created by coherently subtracting $n_1=n_2=1$ photons and performing respective PNR measurement of $n_3=2$ and $n_3=4$, provided the effective squeezing is chosen to satisfy the relation in Eq.~\ref{eq:halfpi_cond} for $\beta_{\ket{0_L}}=\sqrt{3}$ and $\beta_{\ket{1_L}}=1/\sqrt{3}$. In order to satisfy these requirements, the experimental resources must achieve squeezing thresholds of $10.63$ dB to perfectly create the ${|0_L\rangle}$ state and $6.08$ dB to create the ${|1_L\rangle}$ state. Note that for experimental squeezing above these thresholds, the beamsplitter reflectivity must be tuned to decrease the effective squeezing back to these quantities. However, squeezing below these thresholds does not mean that QEC with rotationally-symmetric code words is no longer possible. Lower squeezing means that the lower Fock-components of the code states will contribute more to the superposition, but as long as the coefficients for the two code words are chosen such that they have the equal photon number moments, error correction is still possible as in accordance with Knill-Laflamme condition. In fact, the coefficients deviating from a binomial distribution may prove beneficial in specific circumstances if they are optimized to reduce specific error rates~\cite{Michael2016}.\\
For the present discussion, we limit our focus to the binomial code states of Eq.~\ref{eq:binom_codestates}. We show in Fig.~\ref{fig:fig_probsucc}a how the probability to successfully generated the binomial code states varies with initial resource squeezing. When the squeezing is exactly at the thresholds (10.63 dB for $\ket{0_L}$ and 6.08 dB for $\ket{1_L}$) to generate the desired state, the subtraction beamsplitters must have vanishing reflectivities, resulting in small success probabilities as shown on the left most parts of the plots in Fig.~\ref{fig:fig_probsucc}. For larger available initial squeezing, however, increasing the subtraction reflectivity until the effective squeezing is at the required thresholds boosts the success probability by three orders of magnitude for both code states. 
Note that these larger squeezing values for optimized success probability are within the realm of realistically achievable squeezing~\cite{Vahlbruch2016}.
 \\
In Fig.~\ref{fig:fig_probsucc}b, we display the success probability to generate multi-component 4-fold symmetric states with $k$ Fock-basis components in the superposition given an initial resource squeezing value of $12$ dB as the subtraction beamsplitter reflectivity varies. Generating states with $k$ components in the superposition is a higher order process in the beamsplitter reflectivity as it requires $n_1=n_2 \geq k-1$, so the two component state described by Eq.~\ref{eq:halfpi_symm} is the most common result. It is worth mentioning that the growing number of components in the superposition is  a consequence of extended HOM interference. The post-selection is done for the particular case of $n_1 = n_2$ after the HOM interference, but this can happen in many ways as long as we have $n_1^{(-)}= N-2l, n_2^{(-)} = 2l; l\in[0, N/2]$, where $n_1^{(-)}$ and $n_2^{(-)}$ denotes the number of photons subtracted from modes `$a$' and `$b$', respectively before the HOM inferences at the second balanced beamsplitter, $BS$ 1. Likewise, Fig.~\ref{fig:fig_probsucc}c shows the success probability to generate $k$-component 2-fold-symmetry states with an initial $12$ dB of squeezing.\\
In Fig.~\ref{fig:binom_wig}, we plot the Wigner functions of the generated binomial code words with the optimal effective squeezing of 10.63 dB and 6.08 dB for $\ket{0_L}$ and $\ket{1_L}$, respectively. On the left, the resultant Wigner functions (a) and (c) are for the case of perfect detection ($\eta = 1$) for all three PNR detectors. As expected, the Wigner functions display 4-fold symmetry owing to the photon-number support on every 4th Fock state. In an experimental realization of such state engineering protocol, the photon losses due to propagation though lossy optical components and inefficient PNR detections are inevitable and lead to decoherence of the desired state. \\
Next, we investigate the effects of imperfect detection ($\eta = 0.90$) for $n_3$ measurement for generating both the binomial code words. As evident from the Wigner functions in Fig.~\ref{fig:binom_wig} (b) and (d), the negativity of the Wigner functions is reduced while the phase space symmetry is still preserved. Note that we have considered the ideal detection for $n_1$ and $n_2$ PNR measurements since the deleterious effects of their imperfect detection can be reduced by lowering the reflectivity of the subtracting beamsplitters to ensure that the probability of subtracting more than one photon is negligible. While this prevents the decoherence of the generated code words caused by imperfect detection of $n_1$ and $n_2$, it lowers the overall success rate. Therefore, the PNR detectors with high detection efficiency are desired in such state engineering protocols. 
We then evaluate the state Fidelity, Wigner Negativity, and Wigner Log Negativity (WLN) against the imperfect detection efficiency, $\eta$ for $n_3$ detection. The losses are modeled by setting up a fictitious beamsplittier of transmissivity $\eta$ in the path of $n_3$ photons. Numerical simulations are displayed in Fig.~\ref{fig:wig_negativity}. 
The generated state fidelity, defined as $F(\rho, \sigma) = \big(\text{Tr}[\sqrt{\sqrt{\rho}\sigma\sqrt{\rho}}]\big)^2$, with the ideal binomial code words in Eq.~\ref{eq:binom_codestates} is plotted against the overall detection efficiency in Fig.~\ref{fig:wig_negativity}a. We see that for a given detection efficiency, $|1_L\rangle$ performs better than $|0_L\rangle$, as evident from the larger fidelity difference at lower overall detection efficiency. 
In CV resource theories, the Wigner Negativity (WN), i.e., the minimum value of the negativity of Wigner function, and Wigner Log Negativity (WLN) are the key measures to quantify the non-Gaussian nature of quantum resources which are essential for quantum computational advantages and fault-tolerance~\cite{Quntao}. The WLN is defined as
\begin{equation}
    \text{WLN}:= \text{ln}\bigg[\int|W(q,p)|dqdp\bigg],
\end{equation}
and it immediately follows that a non-negative Wigner function has zero WLN. The WN and WLN are plotted in Fig.~\ref{fig:wig_negativity}b and  Fig.~\ref{fig:wig_negativity}b, respectively. We found that the non-Gaussian nature of the code words is preserved even for the detection efficiency as low as $\eta = 0.5$, as seen from the negative minimum amplitude of the Wigner function, $W(Q,P)_{\text{min}}$ and non-zero value of WLN. It is worth pointing out that the PNR detection efficiency of 98\% is a experimentally feasible with state-of-the-art transition-edge sensors~\cite{fukuda2011titanium}, and methods to obtain $>99\%$ are possible~\cite{gerrits2016superconducting}.  This concludes the realistic implementations of our state engineering scheme for binomial codes.
\subsection{Truncated Cat Code State Generation}
\begin{figure*}
    \centering
    \includegraphics[width = \textwidth]{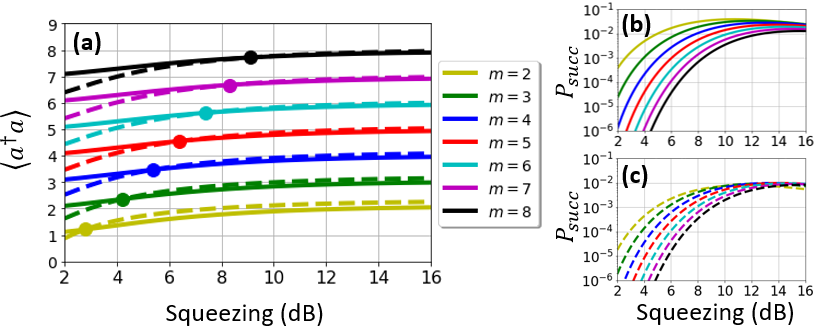}
\caption{(a) Mean photon number of the orthogonal states generated by detecting $[0,1,m]$ (dashed lines) and $[0,2,m]$ (solid lines) photons as initial squeezing increases for a fixed subtraction beamsplitter reflectivity of $10\%$. For each $m$, there is a squeezing value that yields orthogonal states with the same mean photon number (indicated by solid dots) which is necessary to satisfy the Knill-Laflamme criteria. (b) The probability to generate the $[0,1,m]$ or $[1,0,m]$ states, which differ only by an optical phase, and (c) the probability to generate the $[0,2,m]$ or $[2,0,m]$ states.}
    \label{fig:cat_like}
\end{figure*}

In this section, we investigate the feasibility of generating code words having support on higher photon-number. We show that such codes satisfy the Knill-Laflamme condition for faithful error correction against single-photon loss. Consider two events where we perform the detection combinations of $[n_1,n_2,n_3]=[0,1,m]$ and $[n_1,n_2,n_3]=[0,2,m]$ with $m\geq 2$, which we denote as $\ket{\psi_{0,1,m}}$ and $\ket{\psi_{0,2,m}}$, respectively. Using the results from Eqs.~\ref{eq:tms_projected_final} and~\ref{eq:coefAprime}, the former case yields a two component 2-fold symmetric state of the form 
\begin{widetext}
\begin{equation}
    \ket{\psi_{0,1,m}}\propto e^{-i\frac\phi2}(\tanh{R})^{-\frac12}\sqrt{m}\ket{m-1}-e^{i\frac\phi2}(\tanh{R})^{\frac12}\sqrt{m+1}\ket{m+1},
    \label{eq:psi01m}
\end{equation}
while the latter case gives us
\begin{equation}
    \ket{\psi_{0,2,m}}\propto e^{-i\phi}(\tanh{R})^{-1}\frac{\sqrt{m}}{\sqrt{m+1}}\ket{m-2}+2\frac{\sqrt{m+1}}{\sqrt{m-1}}\ket{m}+e^{i\phi}\tanh{R}\frac{\sqrt{m+2}}{\sqrt{m-1}}\ket{m+2}.
    \label{eq:psi02m}
\end{equation}
\end{widetext}
By inspection, it is easy to see that these states are mutually orthogonal where $\ket{\psi_{0,2,m}}$ has parity of $m$ and $\ket{\psi_{0,1,m}}$ has the opposite parity. As seen from the Wigner function plots of the example cases of $m=5$ and $m=7$, these types of states resemble truncated cat-like states with the associated fast interference fringes near the origin in phase-space. If we can arrange for the states to satisfy the Knill-Laflamme condition defined in Eq.~\ref{eq:KL} 
\begin{equation}
    \bra{\psi_{0,1,m}}\hat{a}^\dag \hat{a}\ket{\psi_{0,1,m}}=\bra{\psi_{0,2,m}}\hat{a}^\dag \hat{a}\ket{\psi_{0,2,m}},
\end{equation}
then the two states can be used as code words to protect against single-photon losses. Using the level of squeezing as a tuning experimental knob, this condition can be exactly satisfied. In Fig.~\ref{fig:cat_like}, we show how each pair of states can be made with the same experimental scheme without changing the subtraction beamsplitter reflectivities, and that for the correct squeezing (indicated by the solid dot where the solid and dotted lines cross in Fig.~\ref{fig:cat_like}a ), the two orthogonal states each have the same mean photon number. This means that the scheme can be set with some initial squeezing and beamsplitter parameters and is capable of generating either of the pair of orthogonal states. For higher detected $m$, the two orthogonal states have nearly the same mean photon number for values of squeezing past the exact point as well, leading to slightly larger success probabilities for both $\ket{\psi_{0,1,m}}$ and $\ket{\psi_{0,2,m}}$ as shown in the respective plots in Fig.~\ref{fig:cat_like}b and~\ref{fig:cat_like}c.
The general form of Eq.~\ref{eq:tms_projected_final} can be used to find other pairs of orthogonal truncated cat-like states with more terms in the superposition beyond the simple case of measuring only zero and one or zero and two photons in the subtraction step. However, states created by subtracting larger numbers of photon generally arise less frequently due to decreasing success probabilities with unbalanced beamsplitters, and for arbitrary subtractions of $n_1$ and $n_2$, cat-like state production may not happen for every end PNR detection of $n_3$. Next, we consider how to enlarge the success probability by using multiplexing techniques. 
\begin{figure*}
    \centering
    \includegraphics[width = \textwidth]{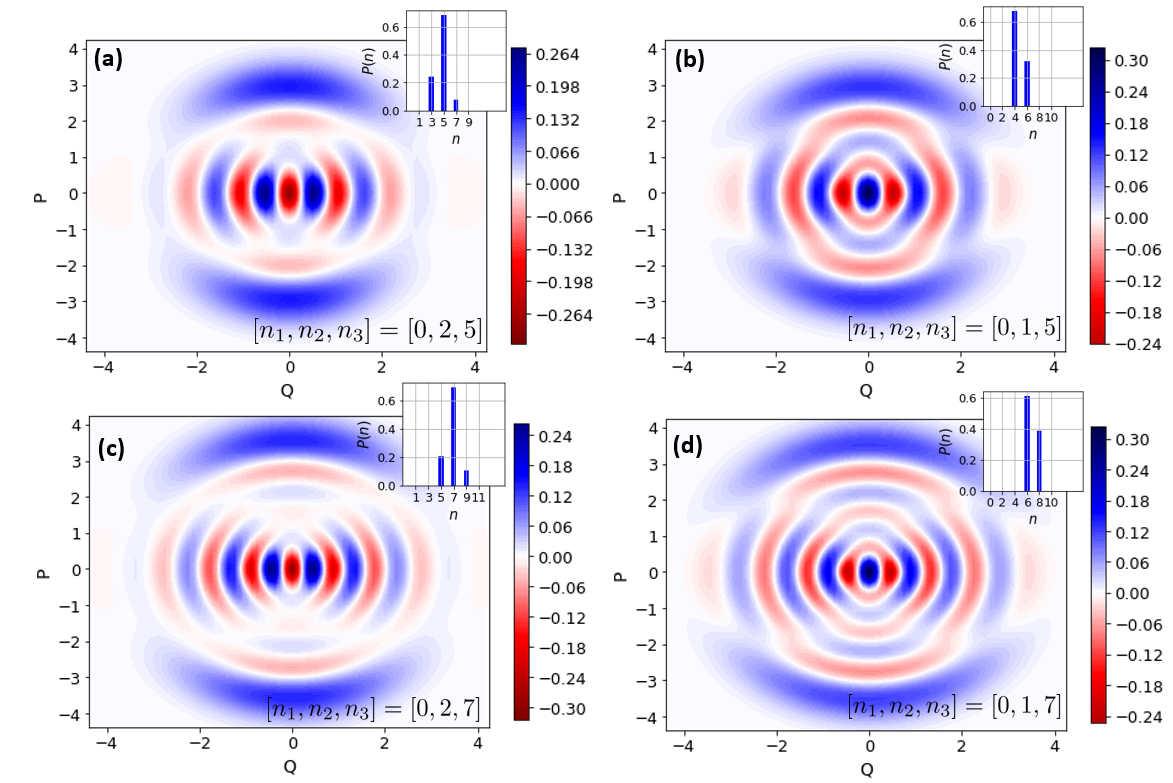}
    \caption{Wigner functions and photon-number distributions of codewords with 2-fold phase space symmetry. (a) and (b) are two orthogonal codewords with mean photon-number $\bar{n} \approx 4.6$ for the measurement outcomes of $[n_1, n_2, n_3] = [0,2,5]$ and $[n_1, n_2, n_3] = [0,1,5]$, respectively. Likewise, (c) and (d) are for $[n_1, n_2, n_3] = [0,2,7]$ and $[n_1, n_2, n_3] = [0,1,7]$ with mean photon number $\bar{n} \approx 6.7$.  
    Insets show the photon-number distributions having support on odd and even photon numbers. The effective squeezing for (a) and (b) is 6.42 dB and 8.30 dB for (c) and (d).}
    \label{fig:cat_like_wig}
\end{figure*}

\subsection{Success probability enhancement by resource multiplexing}
\begin{figure*}
    \centering
    \includegraphics[width = \textwidth]{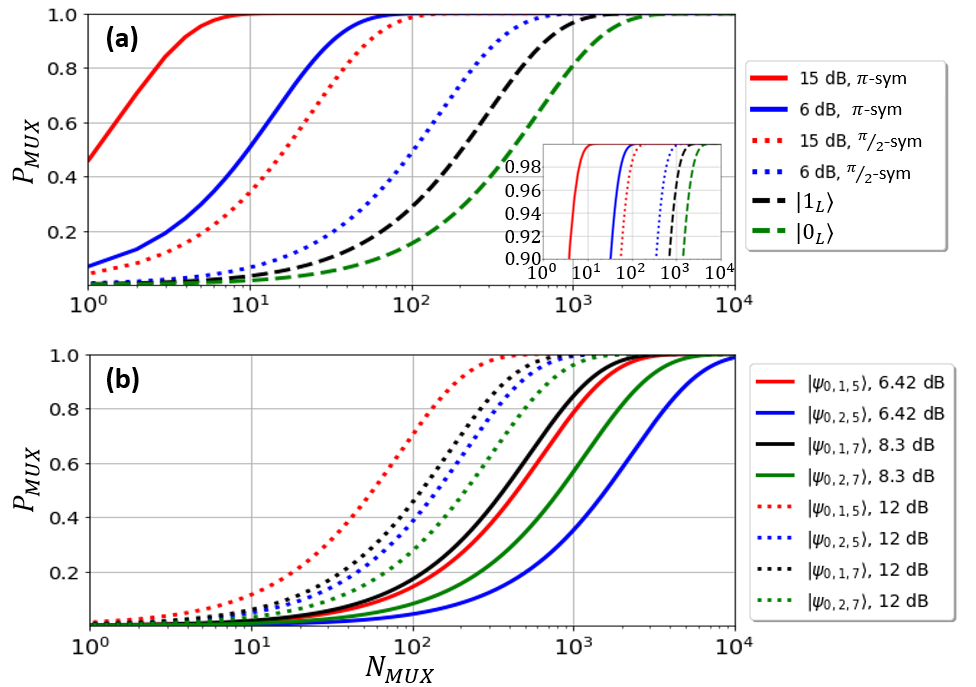}
\caption{Including $N_{MUX}$ multiplexed protocols increases the overall success probability, $P_{MUX}$. (a) Curves are shown for the probability to generate any superposition 2-fold symmetric state using 15 dB and 6 dB initial squeezing (solid red and blue, respectively), any 4-fold symmetric at 15 and 6 dB (dotted red and blue), and the binomial code words $\ket{1_L}$ and $\ket{0_L}$ (dashed black and green) when initial squeezing maximizes the single-shot success probability. (b) The $P_{MUX}$ values are plotted for the truncated cat-like states of the Wigner functions shown in Fig.~\ref{fig:cat_like_wig}. Solid lines correspond to states generated from initial squeezing values that ensure the mean photon number is the same for each orthogonal state in the pair of $\ket{\psi_{0,1,m}}$ and $\ket{\psi_{0,2,m}}$. Dotted lines assumed a flat $12$ dB of initial squeezed, and the mean photon numbers of the pairs of states are now only approximately equal.}
    \label{fig:mux}
\end{figure*}

Thus far, we have considered the generation of bosonic codes using only one setup presented in Fig.~\ref{fig:schematic}. To boost the success probability of generating a particular code one can employ multiplexing schemes, which have been used to enlarge the heralding probability of single-photons~\cite{bonneau2015effect,joshi2018frequency,Kanedaeaaw8586}. For a given success probability $P_{\text{succ}}$ per state generation setup, the probability of generating the desired code with $N_{\text{MUX}}$ multiplexed resources is
\begin{equation}
    P_{\text{MUX}} = 1 - (1-P_{\text{succ}})^{N_{\text{MUX}}}
\end{equation}
For a given success probability tolerance $\delta = 1-P_{\text{MUX}}$, one can estimate the number of multiplexed sources as
\begin{equation}
    {N_{\text{MUX}}} = \bigg{\lceil}{\frac{\text{log}_{10}(\delta)}{\text{log}_{10}(1-P_{\text{succ}})}}\bigg{\rceil},
\end{equation}
where $\lceil . \rceil$ is the ceiling function. In Fig.\ref{fig:mux}(a), we plot the success probability against the number of multiplexed sources for states with both 2-fold (solid line) and 4-fold (dotted line) symmetry at assumed experimental squeezing values of $15$ dB (red) and $6$ dB (blue). The probabilities are calculated using beamsplitter parameters that maximize the single-scheme success probabilities for each case according to Fig.~\ref{fig:all_statetypes}. Fig.~\ref{fig:mux}(a) also shows the the multiplexed probability to generate the $\ket{0_L}$ (dashed green) and $\ket{1_L}$ (dashed black) states where again the other parameters are chosen to maximize individual scheme success probabilities. The figure inset zooms in to show where the multiplexed probabilities exceed 0.90. For respective probability tolerances of $0.1$ and $0.01$, only $N_{MUX}=4$ and 8 schemes are needed to successfully generation a 2-fold symmetric superposition when initial squeezing is $15$ dB.
The second panel of the figure, Fig.~\ref{fig:mux}(b) performs the same multiplexing analysis on the truncated cat-like states whose Wigner function are shown in Fig.~\ref{fig:cat_like_wig}. The solid lines show the $P_{MUX}$ for the cases where the initial squeezing is chosen such that the mean photon number of orthogonal pairs ($\ket{\psi_{0,1,m}} \text{and} \ket{\psi_{0,2,m}}$) is the same. However, the success probability can be significantly increased when larger initial squeezing is available, and orthogonal states still have approximately the same mean number of photons as seen in Fig.~\ref{fig:cat_like}. The dotted lines in Fig.~\ref{fig:mux}(b) show $P_{\text{MUX}}$ when the input squeezing is 12 dB.

\section{Imperfect binomial code words}\label{sec:imper_bin}
In order to engineer the states ${|0_L\rangle}$ and ${|1_L\rangle}$, the final PNR detector must register two and four photons, respectively.  However, when the detector is imperfect and has efficiency $\eta<1$, this leads to mixtures instead of ideal code words. 
We wish to evaluate how well error correction can proceed and try to find a bound on the detector inefficiency beneath which QEC is still possible. In our analysis, we ignore the dark counts of the detector which is a reasonable assumption for superconducting single-photon detectors. Let's begin with the state after coherent photon subtraction where a single photon was measured in each of the subtraction detectors, i.e., $n_1 = n_2 = 1$. The state after photon subtraction is
\begin{equation}
\ket{\psi}\propto(\hat{a}^2 - \hat{b}^2)\ket{0,\mf{R}}_{ab},
\label{eq:TM_loss_startpoint}
\end{equation}
Note that this process can be assumed to be ideal, as slight inefficiencies in the subtraction detectors are less important than the end PNR efficiency since the probability to detect values of $n_1$ and $n_2$ larger than one is an order of magnitude smaller than $n_1=n_2=1$ due to highly unbalanced beamsplitters. It is, however, desirable to have near-unity efficiency in order to subtract photons at higher rates.

For the ${|0_L\rangle}$ code word, performing lossy PNR detection of two photons on mode $`a'$ of Eq.~\ref{eq:TM_loss_startpoint} yields the mixtures \begin{widetext}
\begin{align}
    \rho_{0_L}&=\frac{\text{Tr}_a\left[\Pi_2\ket{\psi}\bra{\psi}\right]}{\text{Tr}\left[\Pi_2\ket{\psi}\bra{\psi}\right]} \nonumber\\
    &\propto\text{Tr}_a\Bigg\{\sum^\infty_{n,n'=0}e^{i\phi(n-n')}(\tanh{\mf{R_{0_L}}})^{n+n'}\Big(\bra{2}_a(a^2-b^2)\ket{n,n}\bra{n',n'}({a^\dag}^2-{b^\dag}^2)\ket{2}_a \\
    &\hspace{12mm}+3(1-\eta)\bra{3}_a(a^2-b^2)\ket{n,n}\bra{n',n'}({a^\dag}^2-{b^\dag}^2)\ket{3}_a+\mf{O}[(1-\eta)^2]\Big)\Bigg\},\\
\end{align}
\end{widetext}
where $\Pi_n = \sum_{m=n} p(n|m)|m\rangle \langle m|$ is the POVM for `m' photons detection from a PNR detector with efficiency $\eta$ and $p(n|m) = {m\choose n} \eta^n (1-\eta)^{m-n}$ is the conditional probability of detecting `n' photons out of `m' photons incidented to the detector. For high efficiency PNR detectors with near unit efficiency, we can neglect orders of $(1-\eta)^2$ and higher. If we assume that the effective squeezing was chosen properly to for the desired state, then we can write the final output as the mixture
\begin{equation}
    \rho_{0_L}=(1-\delta_{0_L}){|0_L\rangle}\bra{0_L}+ \delta_{0_L} \rho_{E_{0_L}},
\end{equation}
where the erroneous component is

\begin{equation}
    \rho_{E_{0_L}}=\frac{1}{8}\left(\sqrt{3}\ket{1}+\sqrt{5}\ket{5}\right)\left(\sqrt{3}\bra{1}+\sqrt{5}\bra{5}\right)
\end{equation}
and the ratio
\begin{equation}
    \frac{\delta_{0_L}}{1-\delta_{0_L}}=3\sqrt{2}(1-\eta).
    \label{eq:deltaup}
\end{equation}

Similarly, one can go through the same procedure for the other code word with a lossly PNR detection of four photons. In this case we have

\begin{align}
    \rho_{1_L}&=(1-\delta_{1_L}){|1_L\rangle}\bra{1_L}+ \delta_{1_L} \rho_{E_{1_L}},\nonumber\\
    \rho_{E_{1_L}}&=\frac{1}{32}\left(5\ket{3}+\sqrt{7}\ket{7}\right)\left(5\bra{3}+\sqrt{7}\bra{7}\right),\nonumber\\
    \frac{\delta_{1_L}}{1-\delta_{1_L}}&=\frac{8\sqrt{2}}{\sqrt{15}}(1-\eta).
\end{align}
The Wigner functions plotted in Fig.~\ref{fig:binom_wig}(b) and~\ref{fig:binom_wig}(d) show the effects the detector inefficiency when $\eta=0.90$. Additionally, Fig.~\ref{fig:wig_negativity} plots the fidelity and Wigner log negativity of the imperfect code states as efficiency deviates from unity.

Now, we wish to examine how the error and recovery process is affected by the imperfections introduced by the non-ideal detector.  Because the chosen code only protects against single photon loss, we need only consider effects up to first order in $\gamma$ from the Kraus operators describing the loss. We will use the approximate code word of $\rho_{0_L}$ as an example and treat the effects of no-jump errors ($E_0$) and single-jump errors ($E_1$) separately.
\subsubsection{No-jump errors}
Expanding in factors of $\gamma$, the Kraus operator for a no-jump event is
\begin{equation}
    E_0=1-\frac12\gamma \hat{a}^\dag \hat{a} + \mf{O}[(\gamma)^2].
\end{equation}
To leading order in $\gamma$, this error acts on $\rho_{0_L}$ to give

\begin{align}
    E_0 \rho_{0_L} E_0^\dag =& (1-\delta_{0_L})E_0{|0_L\rangle}\bra{0_L}E_0^\dag +\delta_{0_L}\rho_{E_{0_L}} \nonumber \\
    & - \tfrac12\delta_{0_L}\gamma \left (\hat{a}^\dag \hat{a}\rho_{E_{0_L}}+\rho_{E_{0_L}}\hat{a}^\dag \hat{a}\right).
    \label{eq:nojump}
\end{align}

The first term in the above expression is the ideal code word transformed by $E_0$ and is proven to be correctable with unitary operations~\cite{Michael2016}. The middle term contains the error from state generation and is orthogonal to the ideal code word. Due to the trace preserving nature of unitary matrices and the orthogonality of the first two components in the mixture, applying the unitary recovery operations will preserve the probability ratio of the error component with the desired corrected component. This means that after the recovery, the error from this term will not grow in magnitude but have the same contribution as the original state preparation error. Thus, if we can neglect the last term in Eq.~\ref{eq:nojump}, then error in the final code word after error and recovery will be the same as that of state generation and not grow in magnitude. In order to neglect this final term, we need $\delta_{0_L}$ to be of the same order as $\gamma$. From Eq.~\ref{eq:deltaup}, we can determine a condition on the PNR detector efficiency in order for the error correction to be feasible up to $\mf{O}(\gamma)$. This condition is
\begin{equation}
    1-\eta \sim \gamma.
\end{equation}
\subsubsection{Single-jump errors}
The expanded single-jump Kraus operator is
\begin{equation}
    E_1=\left(\gamma - \tfrac12(\gamma)^2\right)^{\frac12}a + \mf{O}[(\gamma)^{\frac32}],
\end{equation}
and acts on $\rho_{0_L}$ to yield
\begin{align}
    E_1 \rho_{0_L} E_1^\dag = (\gamma-\tfrac12(\gamma)^2)\Big[&(1-\delta_{0_L})\hat{a}{|0_L\rangle}\bra{0_L}\hat{a}^\dag \nonumber \\
    &+ \delta_{0_L} \hat{a} \rho_{E_{0_L}}\hat{a}^\dag\Big].
\end{align}
The photon loss error is detected by measuring the photon number mod two for our chosen code space, which in this case, is a parity measurement. The error portion of the mixture, $\rho_{E_{0_L}}$, is orthogonal to the ideal state ${|0_L\rangle}\bra{0_L}$, so a measurement after a photon loss might incorrectly yield an even parity result since $\hat{a}\rho_{E_{0_L}}\hat{a}^\dag$ has the same parity as ${|0_L\rangle}\bra{0_L}$. The probability to incorrectly diagnose the loss is thus given by the ratio 
\begin{equation}
    P_{inc}=\frac{\text{Tr}[\delta_{0_L} \hat{a}\rho_{E_{0_L}}\hat{a}^\dag]}{\text{Tr}[(1-\delta_{0_L})\hat{a}{|0_L\rangle}\bra{0_L}\hat{a}^\dag]}\approx \frac{9\delta_{0_L}}{10(1-\delta_{0_L})}.
\end{equation}
Again, since the accuracy of our error correcting code is only up to first order in $\gamma$, if the probability to incorrectly diagnose the error is negligible compared to $\gamma$ then our code still has efficacy. Using Eq.~\ref{eq:deltaup}, this leads to the slightly stronger condition on the detector efficiency of
\begin{equation}
    \eta \gg (1-\gamma).
\end{equation}
This condition states that in order for the QEC to be effective, the losses from detector inefficiency must be considerably smaller than the expected losses the code words will protect against.

\section{Architectural analysis}\label{sec:arch}
In this section, we discuss the experimental feasibility of our proposal with currently available resources. Optical Parametric Oscillators (OPOs) below threshold have been demonstrated to achieve as high as 15 dB of desired squeezing for our scheme~\cite{Schnabel2016}.  
Along with low loss free-space linear optics and fiber coupled highly efficient PNR detectors such as transition-edge sensors (TESs) and superconducting nanowire single photon detector (SNSPDs) ~\cite{Lita2008,Nehra_optica:19,cahall2017multi,fukuda2011titanium}, our proposal can be readily realized. However, this approach requires tight control and active stabilization of the optical cavity lengths and precise free-space alignment, making it challenging to scale for multiplexed experimental setups for boosting the success probability for state generation. Integrated quantum photonics offers a robust, compact, ultra-low loss scalable platform
for building larger quantum circuits including both linear and nonlinear components~\cite{peruzzo2010quantum,metcalf2013multiphoton,najafi2015chip,spring2013boson}. Thanks to the miniaturization offered by photonic integrated circuit (PIC) technology, quantum states can be generated and processed within the same PIC.


The tight transverse field confinement in integrated waveguide structures leads to large effective nonlinearity and alleviates deleterious effects such as gain-induced diffraction that degrades the obtainable squeezing in pulsed pump experiments~\cite{Prem_Kumar_squeezing_2,Prem_Kumar_squeezing,eto2008observation}. Owing to the large nonlinearity, high levels of squeezing can be achieved in the waveguide structures over a bandwidth that is several orders of magnitude larger than OPOs. \\
Over the last couple of decades, the integrated structures based on materials such as Lithium Niobate (LN), Silicon Nitride (SiN) and Silica have been the key enabler for many quantum photonics experiments
~\cite{sanaka2001new,kanter2002squeezing,single_photon_nonlinearity,saglamyurek2011broadband, arrazola2021quantum, dutt2015chip}. While SiN and Silica based integrated architectures have the scalability advantage in nanophotonic integration over LN based diffused waveguides with large cross-sections, they suffers from the weak third-order nonlinearity and unwanted noise, which limits the measured squeezing to only of few dBs~\cite{dutt2015chip,yang2021squeezed}. Recently, the thin-film Lithium Niobate-On-Insulator (LNOI) has lead to several breakthroughs in nanophotonics, owing to sub-wavelength confinement, dispersion engineering and efficient quasi-phasematching using high quality periodic poling, more than an order of magnitude improvements in the nonlinear efficiency compared to traditional LN waveguides have been achieved~\cite{wang2018ultrahigh, Jankowski:20,zhao2020shallow}. Furthermore, ultralow-threshold OPOs~\cite{Tang2021, Safavi2021} and high gain optical parametric amplification and generation have been recently demonstrated~\cite{ledezma2021intense,jankowski2021efficient}. In the quantum domain, thin-film LN waveguide structures with pulsed pump operation offer simple non-resonant and efficient sources for ultra-broadband single-pass traveling-wave phase-sensitive degenerate optical parametric amplifiers (OPAs) for generating squeezed light~\cite{caves1982quantum}. In addition to a strong second-order nonlinear interaction, LN has large electro-optic coefficients and ultra-low loss propagation of $\alpha = 3$ dB/m, which allows for efficient and scalable linear optics implementations within the same chip~\cite{zhang2017monolithic, zhu2021integrated}. Considering the measured nonlinear efficiency and propagation loss in thin-film LN platform, on-chip squeezing levels of $>15$ dB are within the reach of current experimental capabilities~\cite{rao2019actively,jankowski2021dispersion}. On the measurement side, significant progress has been made both in developing low loss chip-to-fiber interconnects~\cite{hu2020high,he2019low} and in monolithic integration of single-photon detectors on the LN thin-film~\cite{sayem2020lithium,colangelo2020superconducting,lomonte2021single}. Our proposal can thus be realized with the current quantum photonics technology.

\section{Conclusions and outlook }\label{sec:conclusions}
In this paper, we explored the coherent photon subtraction from a two-mode squeezed vacuum to generate non-Gaussian quantum states with 2-fold and 4-fold phase space symmetry in the Wigner function, including exact binomial code states and truncated cat-like states. Such states have the desired Wigner function negativity required for quantum advantages in CV quantum information processing. While the proposed method is not a push-button, desired codes can be generated with rates of KHz-MHz by pairing up with state-of-the-art PNR detectors with high count rates. Furthermore, the success probabilities can be enlarged by starting with higher initial resource squeezing and multiplexing schemes, which can be achieved by monolithic integration on the same nanophotonic chip. From architectural analysis we believe that our proposal can be readily realized with already demonstrated squeezing and high quantum efficiency PNR detection. \\
Moreover, our protocol can be extended to increase the rotation symmetry by utilizing PNR based breeding methods. Higher rotation symmetry leads to enlarged spacing in the photon number basis, offering to correct for larger photon loss, gain, and dephasing errors. The truncated finite dimensional Hilbert space makes these codes hardware efficient and amenable for high fidelity unitary operations, thereby opening promising avenues for their applications in some other areas, in addition to fault-tolerant quantum computation, such as quantum metrology and quantum communication in optical quantum information processing. An important future research is to investigate such state engineering protocol for generating higher order codes allowing to correct for higher photon loss, photon gain, and dephasing errors. 

\section*{Acknowledgments}
RN thanks Prof. Joshua Combes for fruitful discussions. ME thanks Chun-Hung Chang and Prof. Nicolas Menicucci for helpful feedback and conversation. RN and AM gratefully acknowledge support from ARO grant no. W911NF-18-1-0285, and NSF grant no. 1846273 and 1918549. The authors wish to thank NTT Research for their financial and technical support. ME and OP gratefully acknowledge support from NSF grant PHY-1708023, Jefferson Laboratory LDRD grant.  We are grateful to developers of QuTip Python Package~\cite{johansson2013qutip}, which was used to perform numerical simulations in this work.

\section{Appendix}\label{sec:appendix}
\subsection{Numerical modeling} \label{appx:num_mod}
Numerical simulations were performed using the QuTip Python Package~\cite{johansson2013qutip}. We define beamsplitter and two-mode squeezing operations as
\begin{align}
    U_{ab} &= e^{\theta(\hat{a}\hat{b}^\dag-\hat{a}^\dag \hat{b})}\\
    S(z)_{ab} &= e^{z\hat{a}^\dag \hat{b}^\dag-z^*\hat{a} \hat{b}}
    \label{eq:num_ops}\\
\end{align}
where $r=\cos{\theta}$ is the beamsplitter reflection coefficient and $z=Re^{i\pi}$ is the complex squeezing parameter. After applying the squeezing operation to vacuum and sending the two modes $a$ and $b$ through beamsplitters, PNR detection is performed using the POVM for a lossy PNR detector detector, which is given as
\begin{equation}
    \Pi_n = \sum_{m=n}^{N_{trunc}} p(n|m)|m\rangle \langle m|,
    \label{eq:lossPOVM}
\end{equation}
where $N_{trunc}=40$ is the size of the Hilbert space used, and $p(n|m) = {m\choose n }\eta ^n (1-\eta)^{m-n}$ is the conditional probability of detecting `n' photons if `m' photons are incident to a PNR detector with the detection efficiency of $\eta$.  

The overall success probability for different states was determined by summing Eq.~\ref{eq:psucc_tot} over all possible values of $n_1,n_2,$ and $n_3$ such that the symmetry condition being tested was satisfied, and the sum was truncated when all remaining probabilities were below $\mf{O}[(10^{-8})]$.
\subsection{Perfect coherent photon subtraction} \label{appx:perfect_sub}
To this point we have made use of the approximation that the subtraction beamsplitters are highly unbalanced so we can assume perfect photon subtraction for subsequent derivations. However, even in a laboratory setting where beamsplitters with non-vanishing reflectivity are used, we can recover the ideal subtraction case with a modification to the value of initial squeezing. In the case where subtraction beamsplitters are given by the operators $U_{ac}=U_{bd}=e^{-\theta(a^\dag b - a b^\dag)}$ and the third beamsplitter to make the subtraction coherent is balanced,$U_{cd}=e^{-\frac\pi4(a^\dag b - a b^\dag)}$, applying these operators to a TMS vacuum yields
\begin{widetext}
\begin{align}
 U_{cd}U_{ac}U_{bd}|0,\mf{R}\rangle_{ab}  =\sum_{n=0}^{\infty}\frac{(e^{i\phi}\tanh{\mf{R}})^n}{n!}\left(ta^\dag+\frac{r}{\sqrt{2}} (c^\dag+d^\dag)\right)^n\left(tb^\dag+\frac{r}{\sqrt{2}} (-c^\dag+d^\dag)\right)^n|0\rangle_{abcd},
\end{align}
\end{widetext}
where $r=\sin{\theta}$ and $t=\cos{\theta}$ are the real reflectivity and transmissivity coefficients of the subtraction beamsplitters. Detecting $n_1$ photons in mode $c$ and $n_2$ photons in mode $d$ leads to the unnormalized output state of
\begin{align}
    |\phi\rangle&=\sum_{n=0}^\infty\sum_{k=0}^{n_1+n_2}B\frac{(t^2\tanh{\mf{R}})^n}{k!(n_1+n_2-k)!}a^{n_1+n_2-k}b^{k}|nn\rangle_{ab} \nonumber \\
    &=\sum_{k=0}^{N}Ba^{N-k}b^k|0,\mf{R'}\rangle_{ab},
\end{align}
where $N\equiv n_1+n_2$ and $B=A_{n_1,n_2,k}$ from Eq.~\ref{eq:coefA}. The above equation is then exactly the same as the perfect coherent subtraction case of Eq.~\ref{eq:tms_projected} with an effective decreased squeezing parameter of
\begin{equation}
    \mf{R}'=\text{tanh}^{-1}{(t^2\tanh{\mf{R}})}.
    \label{eq:sq_eff}
\end{equation}
This shows that the idealized states discuss
ed are in fact equivalent to realistic states with modified squeezing. Furthermore, this fact can be used to increase the overall success probability of a protocol if one has the ability to generate an initial TMS vacuum state with squeezing in excess of what is required to produce a specific target. By beginning with a large squeezing parameter, $\mf{R}$, one can increase the reflectivity of the subtraction beamsplitters to increase the likelihood of successfully subtracting $n_1$ and $n_2$ photons until the effective squeezing has been reduced to the $\mf{R}'$ value required to produce the target.
\subsection{Numerical data}
In Table~\ref{tab:table_sq}, squeezing values at the intersection points in Fig.~\ref{fig:cat_like} are given. 
\begin{table*}\label{tab:table_sq}
\begin{tabularx}{1.0\textwidth}{ |X|X|X|X|X| } 
  \hline
  $m$ & $\langle a^\dag a\rangle$ & Squeezing (dB) & Prob. [0,1,m] &Prob. [0,2,m] 
  \\ 
  \hline 
    2 & 1.21 & 2.78 & $1.39\times 10^{-3}$ & $8.46\times 10^{-5}$ 
    \\
  \hline 
    3 & 2.36 & 4.22 & $1.41\times 10^{-3}$ & $1.82\times 10^{-4}$ 
    \\
  \hline 
    4 & 3.47 & 5.40 & $1.51\times 10^{-3}$ & $3.08\times 10^{-4}$
    \\
  \hline 
    5 & 4.56 & 6.44 & $1.53\times 10^{-3}$ & $4.32\times 10^{-4}$  
    \\
  \hline 
    6 & 5.62 & 7.40 & $1.72\times 10^{-3}$ & $6.28\times 10^{-4}$
    \\
    \hline 
    7 & 6.68 & 8.30 & $1.86\times 10^{-3}$ & $8.34\times 10^{-4}$ 
    \\
    \hline 
    8 & 7.72 & 9.12 & $1.94\times 10^{-3}$ & $1.03\times 10^{-3}$
    \\
  \hline
\end{tabularx}
\caption{Numerical data for the experimental parameters when the mean photon-number is equal for both the code words in Fig.~\ref{fig:cat_like}}
\label{tab:stats}
\end{table*}

\bibliography{Pfister.bib, Thesis.bib}

\end{document}